\documentclass[prl,aps,twocolumn,showpacs,amsmath,amssymb]{revtex4}

\usepackage{graphicx}
\usepackage{dcolumn}
\usepackage{bm}

\begin{document}
\title{Anisotropy in transport properties and susceptibility of single crystals $BaFe_2As_2$  }
\author{X. F. Wang, T. Wu, G. Wu, H. Chen, Y. L. Xie, J. J. Ying, Y. J. Yan,  R. H. Liu }
\author{ X. H. Chen}
\altaffiliation{Corresponding author} \email{chenxh@ustc.edu.cn}
\affiliation{Hefei National Laboratory for Physical Science at
Microscale and Department of Physics, University of Science and
Technology of China, Hefei, Anhui 230026, P. R. China\\ }
\date{\today}

\begin{abstract}
We report that sizable single crystals of $BaFe_2As_2$ have been
grown with the self-flux method. Measurements and anisotropy of
intrinsic transport and magnetic properties from high quality single
crystal are first presented. The resistivity anisotropy
($\rho_c/\rho_{ab}$) is as large as 150 and independent of
temperature, suggesting that transport in the ab-plane and along the
c-axis direction share the same scattering mechanism.  In contrast
to the susceptibility behavior observed in polycrystalline samples,
no Curie-Weiss behavior is observed, and a linear temperature
dependent susceptibility occurs from the spin-density-wave (SDW)
transition temperature, ($T_s$), to 700 K. This result suggests that
strong antiferromagnetic correlations are present well above $T_s$.
A twofold symmetry of susceptibility in the ab-plane indicates a
stripe-like spin structure as observed by neutron scattering. The
resistivity minimum is strongly dependent on the magnetic field,
suggesting that the upturn of the resistivity at low temperatures
should be related to spin fluctuation.
\end{abstract}


\maketitle
\newpage

The discovery of superconductivity in the layered iron arsenic
compounds has generated much interest. Pnictide superconductivity
was initially studied below 26 K in $LaO_{1-x}F_x$FeAs
(x=0.05-0.12)\cite{yoichi} and $T_c$ surpassing 40 K, beyond the
McMillan limitation of 39 K predicted by BCS theory, was obtained in
$RO_{1-x}F_xFeAs$ by replacing La with other trivalent R with
smaller ionic radii \cite{chenxh,chen,ren}. The question naturally
arises as to how the iron-based high $T_c$ superconductors (HTSC)
compare to cuprates. The Fe-based superconductors share some
similarities with cuprates. They both adopt a layered structure. The
doping phase diagram of Fe oxypnictide system is remarkably similar
to that of the cuprates. For both systems, superconductivity emerges
upon doping carriers into an antiferromagnetic parent compound,
while the antiferromagnetic order is suppressed by
doping\cite{yoichi,chenxh,dong}. These similarities are of great
significance because the parent compounds in the two families are
very different. The parent compounds of the pnictides and the
cuprates appear to have different magnetic ground states. The
pnictides appear to be itinerant systems with magnetism arising from
a nesting-induced spin density wave (SDW), in contrast with the
cuprates, which are Mott-Hubbard insulators with antiferromagnetic
ground states. On the other hand, the parent compounds of new
Fe-based superconductors are bad metals and share a similar
electronic structure with five orbitals contributing to a low
density of states at the Fermi
level\cite{dong,singh,lebegue,haule,cao,ma,nekrasov}. This contrasts
to the case of cuprates in which parent compounds are
Mott-insulators with long-range antiferromagnetic ordering.
Therefore, it is very important to study the various behaviors of
the parent compound to compare with the cuprates.

In order to understand high-$T_c$ superconductivity, it is important
to obtain further insight into the differences and similarities of
the pnictide and cuprate HTSCs. A crucial step for study intrinsic,
especially anisotropic, properties is to grow sizable single
crystals. Unfortunately, growth of single crystals of these
materials has proven very difficult. So far, only single crystals
with hundreds of $\mu m$ are reported in the $RFeAsO$
system\cite{zhigadlo,jia}. Superconductivity at 38 K was realized in
$Ba_{1-x}K_xFe_2As_2$ without oxygen\cite{rotter}. As pointed out by
Ni et al., single crystals of such intermetallic compound systems
without oxygen can be grown by the conventional flux
method\cite{ni}. Growth of single crystals with $mm$ size has been
reported with a tin flux, but the single crystal is contaminated
with Sn\cite{ni}. Here we report growth of sizable single crystals
of $BaFe_2As_2$ with the self-flux method to avoid the contamination
by flux. Intrinsic resistivity and susceptibility obtained from
high-quality single crystals are first presented, intrinsic
susceptibility behavior is in contrast to that observed in
polycrystalline samples due to impurities. These intrinsic behaviors
will be helpful for understanding the underlying physics of the
parent compound.

High quality BaFe$_2$As$_2$ single crystals were grown by the
self-flux method. In order to avoid contamination from incorporation
of other elements into the crystals, FeAs was used as the flux. FeAs
was obtained by reacting the mixture of the element in powdered form
in evacuated quartz tubes at 1173 K for 4 hours. Then Ba powder and
FeAs powder were accurately weighed according to the ratio of Ba :
FeAs = 1 : 4, and thoroughly ground. The mixture was loaded into an
alumina crucible and then sealed under vacuum in a quartz tube. The
tube was slowly heated to 973 K at a rate of 120 K/h and kept at 973
K for 200 minutes to allow the reaction of the mixture of Ba and
FeAs. Subsequently, the temperature was raised to 1373 K in 100
minutes, the quartz tube was kept at 1373 K for 1600 minutes, and
then the tube was cooled to 1173 K at a rate of 4 K/h. Finally the
quartz tube was cooled in the furnace after shutting off the power.
The shining plate-like BaFe$_2$As$_2$ crystals were mechanically
cleaved from the flux and obtained for measurements.

\begin{figure}[t]
\includegraphics[width=9cm]{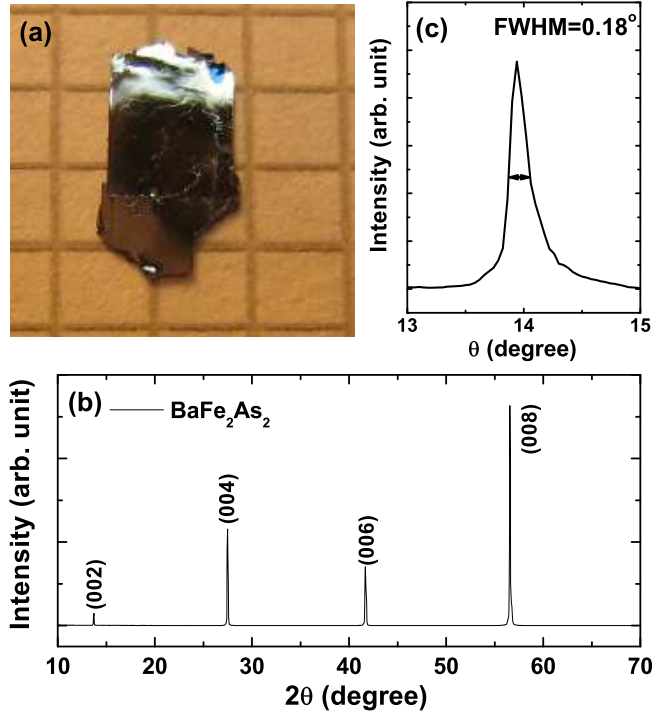}
\caption{ (a): Photograph of a single crystal of $BaFe_2As_2$ on 2
mm grid; (b): Single crystal x-ray diffraction pattern of
$BaFe_2As_2$. Only (00l) diffraction peaks show up, suggesting that
the c-axis is perpendicular to the plane of the plate; (c): Rocking
curve of (004) reflection.\\ }
\end{figure}

Figure 1a shows the picture of a single crystal of $BaFe_2As_2$ on a
2 mm grid. A typical crystal size is 2 x 2 x 0.10 $mm^3$. As shown
in Fig.1a, the linear dimension is as large as 5 mm. All of the
crystals are prone to exfoliation and easily cleaved. Fig.1b shows
the single crystal x-ray diffraction pattern. Only (00l) diffraction
peaks are observed, suggesting that the crystallographic c-axis is
perpendicular to the plane of the plate-like single crystal. The
FWHM in the rocking curve of the (004) reflection is 0.18$^{o}$.
Elemental analysis was performed using energy dispersive x-ray
spectroscopy (EDX). The obtained atomic ratio of Ba:Fe:As is roughly
20.82:40.65:38.53 for all grains; no other phases are detected. The
atomic ratio is consistent with the composition $BaFe_2As_2$ within
instrumental error.

Figure 2 shows the temperature dependence of in-plane and
out-of-plane resistivity measured with PPMS (Quantum Design Inc.).
Both in-plane and out-of-plane resistivity show similar temperature
dependent behavior. The in-plane and out-of-plane resistivities show
metallic behavior above 138 K and a steep decrease at 138 K. This
behavior is consistent with that of polycrystalline
samples\cite{rotter1,wu}, but the residual resistivity ratio (RRR)
of about 2 in single crystal is less than 4 in polycrystalline
sample\cite{wu}. It could arise from a metallic impurity
contribution (such as Fe) in the polycrystalline samples. The
results observed here for single crystal $BaFe_2As_2$ are in sharp
contrast to those from single crystals $BaFe_2As_2$ grown by Ni et
al.\cite{ni}. This contrasting behavior could arise from the
contamination of Sn from the tin flux in Ref.15. It suggests that
the use of the self-flux (FeAs) technique is very important to
obtain single crystals without contamination. It further indicates
that we are able to probe the intrinsic properties of the
superconductor in our clean single crystal grown by the self-flux
method. The inset shows the resistivity anisotropy
($\rho_c/\rho_{ab}$). $\rho_c/\rho_{ab}$ is as large as 150,
indicating that the system is quasi-two dimensional. The near
temperature independence of the ratio $\rho_c/\rho_{ab}$ suggests
that in-plane and out-of-plane transports share the same scattering
mechanism. It should be addressed that all results discussed here
are well reproducible. We measured more than ten crystals to check
the results reproducible from crystal to crystal. In order to make
sure that out-of-plane resistivity measured is intrinsic, and is not
dominated by microcracks between the layers, single crystals used
for measurements are characterized by rocking curve and the rocking
curve just shows the  sharp single peak (no splitting) for all
single crystals as shown in Fig.1c.

\begin{figure}[t]
\includegraphics[width=9cm]{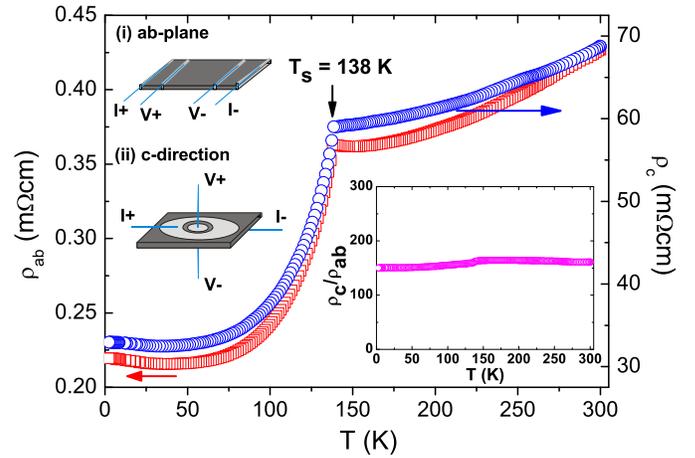}
\caption{Temperature dependence of in-plane and out-of-plane
resistivity ( $\rho_{ab}(T)$ (squares) and $\rho_c(T)$ (circles) )
for single crystal $BaFe_2As_2$. The inset shows that the
resistivity anisotropy ($\rho_c/\rho_{ab}$).
$\rho_c/\rho_{ab}$ is independent of temperature, indicating that
transport in the ab plane and along the c-axis direction share the same
scattering mechanism. \\}
\end{figure}

The temperature dependence of the susceptibility ($\chi$), measured
with SQUID (Quantum Design Inc.) under the magnetic field (H) of 6.5
T applied in the ab-plane and parallel to the c-axis, is shown in
Fig.3a. It should be pointed out that an anisotropy between
H$\parallel$ab plane and H$\parallel$c-axis is observed. As shown in
Fig.3a, the susceptibility decreases monotonically with decreasing
temperature and shows a linear temperature dependence above 136 K.
At 136 K corresponding to the anomaly in resistivity shown in Fig.2,
the susceptibility shows a rapid decrease which should be ascribed
to the occurrence of an antiferromagnetic SDW. Below 136 K the
susceptibility decreases more strongly than T-linear dependence. A
slight increase in susceptibility is observed at low temperatures.
These behaviors are in contrast to that reported in polycrystalline
sample $LaOFeAs$ and $BaFe_2As_2$\cite{rotter1,yoichi}. In
polycrystalline samples, a strong Curie-Weiss tail is always
observed at lower temperatures. Such a strong Curie-Weiss tail
arises from impurities. Interestingly, the linear-T susceptibility
has been widely observed in undoped/highly underdoped
$La_{2-x}Sr_xCuO_4$\cite{nakano}, in $Na_{0.5}CoO_2$\cite{foo} and
in chromium and its alloys\cite{fawcett}. The linear-T
susceptibility occurs before the antiferromagnetic transition for
both $La_{2-x}Sr_xCuO_4$ and $Na_{0.5}CoO_2$, and above the
antiferromagnetic SDW transition for chromium and its alloys, just
like in the pnictides. Such linear-T susceptibility indicates a
strong antiferromagnetic correlation. Recently, Zhang et al. give a
theoretical discussion on the linear-T susceptibility\cite{zhang}.
The inset of Fig.3a shows a linear-T susceptibility up to 700 K,
similar to the case of highly underdoped $La_{2-x}Sr_xCuO_4$, in
which $\chi$ increases linearly with temperature up to higher than
600 K\cite{nakano}. In addition, the value of the susceptibility at
400 K is about 1x$10^{-3}$ emu/mole for $BaFe_2As_2$, while it is
about 0.17x$10^{-3}$ emu/mole for pure Cr\cite{fawcett}. The
susceptibility of $BaFe_2As_2$ is much larger than those of the $3d$
electron based superconductors, e.g., about $10^{-4}$ emu/mole for
$La_{2-x}Sr_xCuO_4$\cite{tagaki}. The linear-T susceptibility with
large magnetization implies the coexistence of local moment and
itinerant SDW in the pnictide superconductors. The coexistence of
local moment and itinerant electrons has been observed in
$Na_xCoO_2$\cite{balicas}.

The susceptibility is measured at 4 K under the magnetic field
$H=6.5$T rotated within the ab plane. As shown in Fig.3b, a twofold
symmetry in susceptibility is observed, and a jump in susceptibility
happens at certain angles. It should be emphasized that the twofold
symmetry disappears above 136 K, indicating that the twofold
symmetry is likely related to SDW ordering. These results indicate
that an easy axis exists in the ab plane and a stripe-like spin
ordering occurs as observed by neutron scattering\cite{cruz,huang},
and the spin prefers to one direction. The jump in susceptibility
arises from that spin orientation along the easy axis is strongly
pinned, and shows up when the pinning force cannot overcome
anisotropy energy. Although the spin prefers one direction and there
exists anisotropy energy due to pinning, the anisotropy of
susceptibility in the ab plane is small, just 1.14. It should be
pointed out that the isothermal M-H curves have been measured under
the magnetic field (H) applied along c-axis and within ab-plane from
2 K to 300 K. All results show that M is proportional to H in the
whole temperature range from 2 to 300 K for different orientations
of H within ab-plane and along c-axis. It suggests that the
susceptibilities shown in Fig.3a and 3b are independent of H.

\begin{figure}[h]s
\includegraphics[width=9cm]{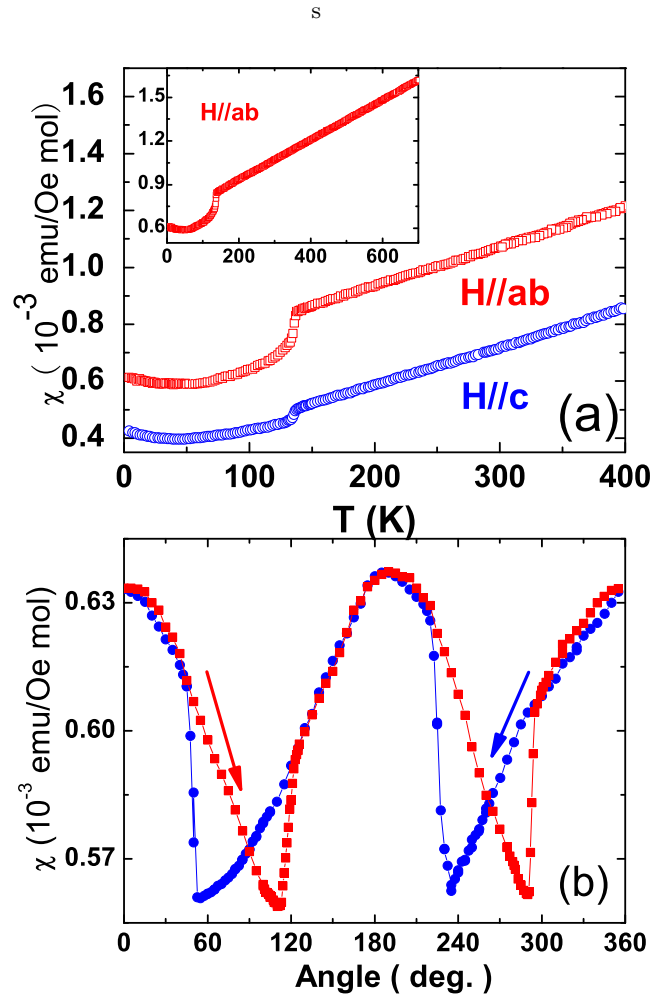}
\caption{Temperature dependence of susceptibility measured under
H=6.5 T applied along the ab-plane and along the c-axis, respectively.
The inset shows susceptibility up to 700 K. (b): Susceptibility at 4 K
as a function of magnetic field angle (H=6.5T) within ab
plane. A twofold symmetry in susceptibility
under rotation in the ab plane indicates a stripe-like
magnetic structure, as observed by neutron scattering in
$LaFeAsO$\cite{cruz}.
\\}
\end{figure}

The temperature dependence of both in-plane and out-of-plane
resistivity is shown in Fig.4 under $H=0$, $H=6.5$T within the ab
plane, and H=6.5T along the c-axis, respectively. Almost no in-plane
and out-of-plane magnetoresistivity is observed above the SDW
transition. Below 100 K, apparent in-plane and out-of-plane
magnetoresistivity appears. Both in-plane and out-of-plane
magnetoresistivity are as large as $10\%$ at 5 K under H=6.5 T along
the c-axis. The effect of H on in-plane resistivity is much the same
with H applied within the ab plane or along the c-axis. The inset
shows results of $\rho_{ab}$ and $\rho_c$ vs. $T$ in the low
temperature range. Both zero-field in-plane and out-of-plane
resistivity show a minimum at the same temperature of $\sim 35$ K.
Below 35 K, the in-plane and out-of-plane resistivities follow a
$log(1/T)$ behavior and deviate from $log(1/T)$ behavior below $\sim
12$ K and saturate at low temperatures. This behavior could arise
from the Kondo effect. Applying a magnetic field leads to a shift of
\begin{figure}[h]
\includegraphics[width=9cm]{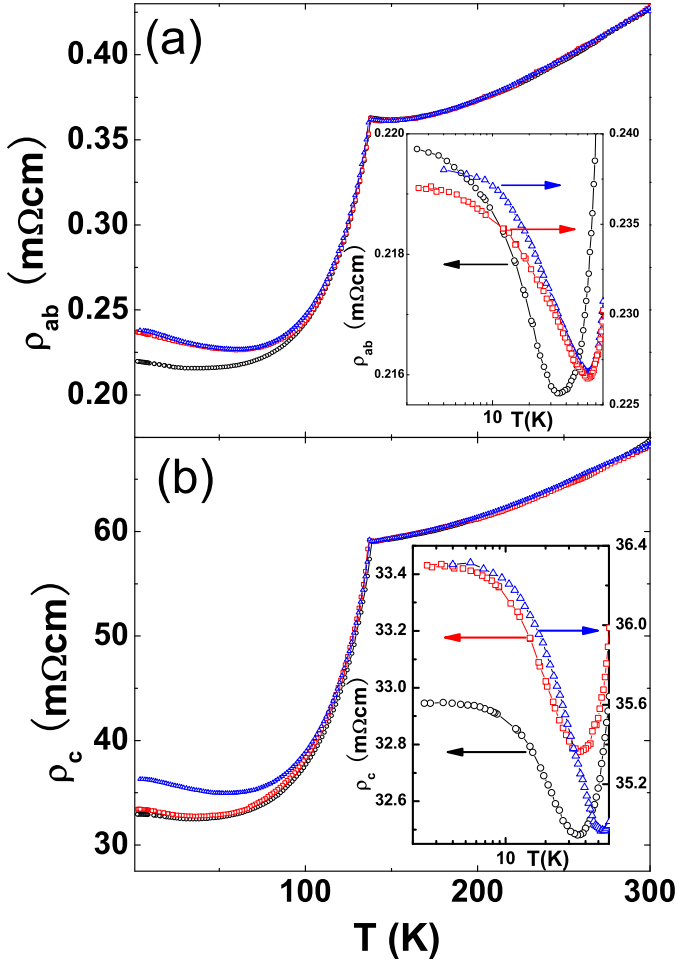}
\caption{Temperature dependence of in-plane and out-of-plane
resistivity under magnetic field (H) of 0 and 6.5 Tesla with H
$\parallel$ to the ab plane and along the c-axis, respectively.
Zero-field resistivity (circles); H $\parallel$ ab plane (squares);
and H $\parallel$ c-axis direction (triangles). Insets show the plot
of low temperature $\rho_{ab}$ and $\rho_c$ vs. T in log scale.\\}
\end{figure}
the resistivity minimum to higher temperature. The temperature of
the resistivity minimum is the same under the applied magnetic field
within the ab plane and along the c-axis.  The temperature of the
resistivity minimum shifts to 60 K and 53 K for in-plane and
out-of-plane applied field of $H=6.5$ T, respectively. Strong
dependence of the resistivity minimum on the applied magnetic field
suggests spin fluctuations or the Kondo effect, instead of
localization in the pnictides. The Kondo effect, the most likely
possible mechanism, arises from magnetic impurity scattering from
defects at Fe sites. The spin fluctuations can be enhanced by H as
observed in cuprates.

In summary, we systematically study the anisotropy of resistivity
and susceptibility in high-quality single crystal $BaFe_2As_2$. The
results indicate that the system is quasi-two dimensional. A linear
temperature dependent susceptibility is observed well above
$T_s=136$ K up to about 700 K, indicating strong antiferromagnetic
correlations well above $T_s$. Notably, the susceptibility behavior
we observe is in contrast to all observations of polycrystalline
samples, due to impurity effects. These intrinsic behaviors are
helpful to understand the underlying physics of the parent compound.
The strongly magnetic field-dependent resistivity minimum is similar
to those of underdoped cuprates. These results will help us to
compare the pnictides with the cuprates and to understand the
high-$T_c$superconducting mechanism.

\vspace*{2mm} {\bf Acknowledgment:} We thank S. Y. Li and D. L. Feng
for valuable discussion and suggestion. This work is supported by
the Nature Science Foundation of China and by the Ministry of
Science and Technology of China (973 project No: 2006CB601001) and
by National Basic Research Program of China (2006CB922005).

\end{document}